# Cutoff-Free Traveling Wave NMR[†]

**Joel A. Tang[1], Graham C. Wiggins[2], Daniel K. Sodickson[2] and Alexej Jerschow[1,2]***

[1] Department of Chemistry, New York University, New York, NY, 10003.

[2] Center for Biomedical Imaging, Department of Radiology, New York University Langone Medical Center. New York, NY, 10016.

* To whom correspondence should be addressed.
Phone: 212-998-8451 Fax: 212-260-7905 E-mail: alexej.jerschow@nyu.edu.

**Abstract**

Recently, the concept of traveling-wave NMR/MRI was introduced by Brunner et al. (*Nature* **457**, 994-992 (2009)), who demonstrated MR images acquired using radio frequency (RF) waves propagating down the bore of an MR scanner. One of the significant limitations of this approach is that each bore has a specific cutoff frequency, which can be higher than most Larmor frequencies of at the magnetic field strengths commonly in use for MR imaging and spectroscopy today. We overcome this limitation by using a central conductor in the waveguide and thereby converting it to a transmission line (TL), which has no cutoff frequency. Broadband propagation of waves through the sample thus becomes possible. NMR spectra and images with such an arrangement are presented and genuine traveling wave behavior is demonstrated. In addition to facilitating NMR spectroscopy and imaging in smaller bores via traveling waves, this approach also allows one to perform multinuclear traveling wave experiments (an example of which is shown), and to study otherwise difficult-to-access samples in unusual geometries.

---

**Introduction**

NMR spectroscopy and MRI are usually thought of as near-field phenomena due to the small sample dimensions compared to the typical wavelengths used (which may be on the order of meters). Although the advent of high field MR scanners has blurred the boundaries, requiring full-wave electrodynamic analysis to account for the observed patterns of signal transmission and reception, until recently, radio frequency (RF) coil design has generally focused on the creation of desirable static or standing-wave field patterns rather than on the generation of traveling waves per se. In a recent report,[1] the NMR phenomenon and MR images were demonstrated in a setting that was deliberately designed to allow the use of traveling waves both in RF irradiation and in the detection of nuclear magnetization signals. In this work, an antenna for transmitting and receiving rf signals was placed at the end of the gradient RF shield of an MR scanner. The magnet bore acted as a waveguide, allowing the waves to propagate from the antenna through the bore and partly through the sample (in this case an imaging phantom or a human body), and also to propagate back through the bore to the antenna for signal reception. This method was fittingly named Traveling Wave NMR (TWNMR). The original proof-of-principle study has led to a number of follow-up studies involving imaging of the human head[2-4] and body,[5-8] and to the development of alternative antennas, waveguide structures, and applications.[3,9-13] Whereas the presence of large quantities of dielectric tissue within the magnet bore clearly distorts and complicates the simple picture of waveguide mode propagation,[4,6,8,14] the general concept of TWNMR has the potential to open up new degrees of freedom for MR excitation and reception.

In the TWNMR study of Brunner et al,[1] the waveguide nature of the magnet bore placed a limitation on the propagation frequencies that could be supported as traveling waves. Frequencies below the waveguide cutoff frequency (303 MHz with a bore diameter of 58 cm) would not propagate. This cutoff frequency was just slightly above the operating frequency of the magnet (298 MHz). One may lower the cutoff frequency by placing dielectrics in the bore and thereby allow $^1$H measurements to be performed. The cutoff behavior, however, nevertheless significantly limits the types of measurements that can be performed presents great challenges for applying the technique at lower field strengths.

For typical NMR spectrometer magnets, the operating frequencies can be much higher, but the bore dimensions are also smaller. Even small animal imagers will have smaller dimensions and therefore higher cutoff frequencies than human scanners, essentially making the traveling wave approach impossible. With typical bore dimensions of 89 mm (wide-bore magnet), the cutoff frequency is 1.97 GHz, and with 51 mm (standard bore) it is 3.45 GHz, both of which are too large to be useful. One could reduce the cutoff frequency by filling the bore completely with dielectric material (e.g. $H_2O$ with $\varepsilon=78.4$ @ $25^o$C), thereby reducing the cutoff frequency by a factor ~8.9. Apart from the inconvenience of filling the bore completely with dielectric material, the cutoff frequency would still be relatively high (~220 MHz and 380 MHz, respectively for the two bore dimensions).

Here, we describe an alternate approach for realizing TWNMR. By turning the waveguide into a transmission line (TL), one can avoid the costly cutoff behavior. TLs support transverse electric and magnetic (TEM) modes, the same types of modes as in free space, which do not have a cutoff frequency.[15] The form of a coaxial TL, where an inner conductor follows the axis of the outer cylinder, is particularly convenient to implement, although many different designs with additional conductors will allow TEM-like modes to propagate without cutoff. A TL can also be matched relatively easily to the transmission circuit, and it can also be terminated with a matching impedance, thus minimizing the reflected power that might occur in an open magnet bore.

We show traveling wave excitation and detection of NMR signals from a sample within a transmission line, at cross-sectional dimensions which would prohibit traveling wave behavior in a cylindrical waveguide, and also demonstrate the broadband nature in a heteronuclear experiment. Although transmission lines have been used previously for probe tuning purposes,[16-23] in which case they are simply used for their capacitive or inductive properties, or for the design of homogeneous resonators,[24,25] where similar to our work, the sample is placed within an expanded region of the transmission line. In the latter, the goal was specifically to construct a resonator with standing wave operation in mind. As a result, the traveling wave mode regime is specifically excluded, and likewise no broad banded tuning behavior is achieved that would allow multinuclear experiments to be performed. A further similarity of our approach may be noted with toroid cavity resonators,[26-28] where simply the geometrical implications of having a coaxial resonator arrangement are exploited, or with the slotted tube resonator design,[29,30] where the field-shaping properties of concentric conductors are used to obtain a relatively homogeneous field distribution. As such, this again does not represent a traveling wave mode of operation. By contrast, we show below that signals and images can be obtained in traveling mode operation well below the cutoff frequency of the cross-sectional dimension of the TL, and include a demonstration of heteronuclear experiments.

**Experimental**

*Sample Preparation.* Mineral Oil and diethyl phosphite were purchased from Sigma-Aldrich and were used without further purification.

*Fabrication of Transmission Line.* Coaxial cables, terminators, and connectors were purchased from Pasternack Enterprises (www.pasternack.com).

A RG223 coaxial cable was cut to a length such that the cable line extends through and beyond the bore of the magnet. The total length of the cable (measured from the amplifier of the NMR spectrometer to the end of the cable) was selected to be a multiple of the wavelength ($\lambda$) at the Larmor frequency ($\nu$) of the observed nucleus to reduce any reflections made with the traveling-wave mode of operation. For the present study $\nu(^1H) = 400$ MHz corresponding to $\lambda = 51.74$ cm and the total cable length was approximately 21 $\lambda/2$.

The sample was placed within the transmission line by cutting away the outer conductor and insulator to expose the inner conductor of the coaxial cable. The sample section was placed at a distance of $3\lambda/2$ away from the end of the transmission line to ensure that the sample position could be varied without bringing the cable end into the magnet. A glass tube of length 4.08 cm and a 5.46 mm i.d. was placed over the exposed inner conductor and sealed using Teflon™ tape and a cyanoacrylate-based adhesive (Super-Glue®) at both ends. The glass tube was covered with conductive copper foil (25.4 mm wide and 0.066 mm thick; www.grainger.com) which was soldered to the outer conductor of the cable (Figure 1). The sample was injected/removed using a syringe at the top of the chamber through the Teflon™ seal.

*NMR Experiments.* All experiments were performed on a Bruker Avance 9.4T wide vertical bore NMR spectrometer. TL experiments were compared to static (non-spinning) experiments performed using a 4.0 mm HX MAS NMR probe. Initial test experiments were conducted on a sample of mineral oil because it has a dielectric constant ($\varepsilon$) of 2.1 which was close to that of the inner insulator of the coaxial cable ($\varepsilon$(polyethylene) = 2.3), keeping the impedance of the line approximately constant across the junctures of the sample chamber. Other materials could be matched using an impedance transformation circuit, or by changing the geometry of the TL.

The TL was attached to the NMR spectrometer via the preamplifier and passes through the center of the magnet bore, thus, the static magnetic field ran parallel to the conductors of the TL (Figure 1). The TL circuit was either attenuated using a 50 Ω attenuator adaptor, shorted with a coaxial short connector or left open at the end of the cable (or terminated end).

Single pulse $^1$H NMR experimental parameters with the 4.0 mm HX MAS NMR probe entailed a 90° pulse width ($\tau_{\pi/2}$) of 5.88 μs. With the TL setup, the pulse width was set to 212, 100 and 500 μs when the TL was attenuated, open and shorted, respectively, which corresponded to the responses with the largest signal. One scan was used to acquire the signal with both setups.

*NMR Imaging Experiments.* $^1$H 2D imaging experiments with the TL setup were performed using a projection reconstruction method.[31] The gradient shim coils of the 9.4T magnet were employed for the source of the gradients where the strengths along the X and Y directions were 0.6 mT/m. 40 gradient settings were utilized to produce different gradient direction angles in the plane of the image and are manipulated using a back-projection algorithm based on the inverse Radon transform to reconstruct the image.[32] A spin-echo sequence was used for each angle. The echo delay was set to 0.41 s so that all signals with fast relaxation (i.e. alkane chain backbone) would dephase and only one site (i.e. methyl groups) would be used as the imaging agent. 8192 data points covering a spectral range of 5000 Hz were used and 16 scans were collected for averaging.

*Multinuclear Experiment.* 2D $^1$H-X NMR experiments were performed on a sample of diethyl phosphate (ε = 23). It should be noted that, although the significantly different dielectric constant would create some reflection with the travelling wave (hence reducing signal intensity), the dimensions and length of the TL were not altered since a sufficient amount of signal was observable to provide evidence for the experimental setup.

For the TL setup, a T-connector combined the X and $^1$H channels to the single TL that is placed along the center of the magnet bore. High power band pass filters were placed in line with each channel to ensure that only the respective frequencies are passed. The line was terminated with a 50 Ω attenuator. $\tau_{\pi/2}$ and $\tau_{\pi}$ were set to 400 and 800 μs for $^1$H and $\tau_{\pi/2}$ = 275 μs for $^{31}$P.

For the $^1$H-$^{31}$P HMQC experiments with the MAS probe, $^1$H and $^{31}$P $\tau_{\pi/2}$ values were set to 100 and 23.75 μs, respectively. The duration of the π pulse was misset to 360 μs to mimic the action of imperfect pulses of the TL since the power decreases radially from the center conductor of the TL, causing the flip angles to be non-uniform across the sample. $^1$H decoupling was not used during acquisition. The phase cycling of the HMQC sequence was adjusted to compensate for the distribution of flip angles in the TL. ($\varphi_1$ = x, $\varphi_2$ = (x)$_4$ (y)$_4$ (-x)$_4$ (-y)$_4$, $\varphi_3$ = x, -x, $\varphi_4$ = x, x, -x, -x, $\varphi_R$ = x, -x, -x, x, -x, x, x, -x).

For both experiments, the spectrum consisted of 4096 × 128 complex points and 64 scans were collected for each $t_1$ time point with a recycle delay of 4 s. The spectral widths of $F_1$ and $F_2$ were 4771 and 7508 Hz, respectively.

**Results and Discussion**

Figure 2 compares the NMR spectra acquired using a 4.0 mm magic-angle-spinning (MAS) probe with the one using the assembled TL. Similar experimental parameters were used for both measurements except for the sample volumes; V(TL) = 725 μL, V(Rotor) = 75 μL. The pulse duration was longer for the TL to obtain the maximum signal. Using the TL, an almost identical spectrum was acquired as the one obtained with the MAS probe. The signal intensity was 67.4% smaller in the TL than in the NMR probe, but use of the TL still resulted in a reasonably high resolution spectrum. The sensitivity of the TL was lower by a factor 28. Of this loss, approximately 20% was due to the non-uniform RF excitation in this geometry (*vide infra*). Some further loss was also due to the non-uniform receive profile. As with TWNMR, a lower efficiency is expected, in general, because the waveguide or the TL are essentially untuned.[1]

The distribution of RF field amplitudes in the sample was tested by performing a series of experiments with increasing pulse widths. A Fourier transform with respect to the pulse width gives the flip angle distribution. Figure 3 shows such a flip-angle distribution for this sample in the form of a histogram of numbers of spins vs. flip angles. In a coaxial cable, the magnetic field depends inversely on the distance from the center. Together with a square weighting due to the change in circumference with distance, the theoretical curve of Figure 3 is obtained, which agrees well with the measured one, except near the surfaces of the inner and outer conductors, where some non-uniformity is expected. If one chooses a pulse flip angle that maximizes the observed signal, one obtains from this RF distribution a reduction in signal to 80.7% as compared to a uniform excitation profile.

In order to verify the traveling wave nature of the measurement process, the experiments were also performed with open, shorted, and matched termination ends of the TL. The appearance of standing waves in the open and shorted configurations was examined by measuring the $^1$H NMR signal intensities of the mineral oil with different cable lengths. Extension cables ranging between $\lambda/2$ and $3\lambda/2$ at intervals of $\lambda/8$ were added before and/or after the sample position in order to test reflection behavior.

Figure 4a illustrates the different cable lengths and configurations used. Figure 4b shows the integrated area of the signal as a function of wavelength as cable is added before the sample region. A small modulation in the signal intensity was observed as a function of additional line length when the line was terminated with a matched attenuator. With an open termination, the amplitude of the signal increased by a factor of approximately 1.6 illustrating that the sample is positioned near a standing wave maximum, and when terminated by a short, the signal dropped to near zero. The fact that only slight oscillations occur as a function of length in the matched termination configuration, indicates that the sample region is matched very well to the cable. If the TL were not matched, the waves would change (amplitude and wavelength) at the boundary of the coaxial cable and the sample, creating some interference and may result in signal loss.

Traveling wave character was tested by checking for absence of standing waves by variation of cable length after the sample with various terminations (Figure 4c). In the attenuator-terminated matched configuration, little variation is seen as opposed to the open and shorted configurations. The phase difference between the open and shorted configurations is 89.6°. Shorted and open terminations therefore produce behavior akin to those of regular TLs. These results indicate that when the line is attenuator terminated, almost no reflection occurs and the TEM waves are indeed traveling through the line, through the sample, and power is almost entirely absorbed in the attenuator. The results from the open and shorted configuration also indicate that the waves behave as if the sample were not present, meaning that the match between sample and cable is very good.

Cutoff-free TWNMR 2D images were acquired using a projection-reconstruction method with matched, open, and short terminations with a $4\lambda$ cable length ($\lambda/2$ cable extension added) after the sample (Figure 5). This cable length was chosen to maximize the signal difference between the open and the shorted configurations. A circular image was obtained in all cases, which replicates the longitudinal cross section of the sample holder. The low signal at the center of the image corresponds to the position of the inner conductor in the center of the tube. The signal intensity of the images also varied when using an open or a short circuit, as expected when using a 4λ cable length after the sample (*vide supra*). The maximum signal occurs at some radial distance away from the central conductor, which is a consequence of the flip angle distribution, where the maximum transverse magnetization occurs away from the center and leads to the observed 'doughnut' shape (The flip angle closest to the inner conductor is larger than $\pi/2$ for the first pulse. Additional distortions are introduced by the inversion pulse.).

The cutoff-free TWNMR approach as implemented in this work will also enable experiments virtually at any field strength and with a variety of sample dimensions. As a consequence, one can also perform these experiments with lower-γ nuclei, such as $^{23}$Na, $^{29}$Si, $^{31}$P, $^{19}$F, and one could easily perform multinuclear experiments without special tuning considerations (Figure 6a). As an example, $^{1}$H-$^{31}$P HMQC NMR experiments (Figure 6b) were performed using the 4mm HX MAS NMR probe and a double resonance TL setup on a sample of diethyl phosphite ($\varepsilon = 23$). The spectrum obtained with the TL setup shows proper single bond H-P correlation and the correct $J(^{1}\text{H}-^{31}\text{P})$ coupling in the direct dimension.

As in ideal TWNMR, where the transfer between the media is unhindered, about half the signal would be lost, because it divides into a forward and a backward propagating wave. These losses could be recovered by placing a transmit/receive switch at the end of the TL in combination with a signal combiner or an additional receiver.[33,34] The TL of the sample chamber need not be coaxial, of course, but could be of any arrangement that would fit the geometrical or tuning requirements. Figure 7 shows a possible stripline arrangement. The construction would be similar as described above, however the sample would be placed between two parallel plates that can be in any shape or orientation. The advantage of using a stripline configuration is that the $B_1$ field between the plates is highly homogeneous.[35,36]

In view of possible MRI experiments, one could place a conductor in the center of the bore of an MRI scanner, or one could line the bore of the magnet with a pair of strips for better rf homogeneity. Alternatively, one could line the patient table with conducting material. Such an arrangement would also lead to a more homogeneous rf field distribution than a coaxial arrangement. An antenna could likewise be used to couple to the bore similarly to the waveguide experiments, with the difference being that now also modes with frequencies below the cutoff of a corresponding waveguide would propagate. When the bore is smaller, the bore TL could be matched with the aid of a matching circuit or an impedance transformer. Further fields of application may include the measurement of many samples simultaneously, which would be placed along the inside of a TL or a stripline, perhaps on a conveyer belt. Analysis of fluid flow could be performed over extended lengths.

**Conclusions**

In this work it is demonstrated that traveling RF waves through the sample can be used to excite nuclear spins and detect NMR signals without the limitation of a cutoff frequency. Such an approach may allow for the remote detection of NMR/MRI signals in unusual or small sample geometries. A further opportunity is the simultaneous measurement of a variety of samples / nuclei at different positions along a TL, without the need of specific tuning.

In general, all TWNMR implementations with large dielectric and conducting samples (e.g. human bodies) face the limitations of field attenuation in the sample and additional reflections, which reduce the traveling wave character of the measurements. However, one of the greatest advantages of TWNMR is that it is very simple to implement. For example, building very large volume coils is non-trivial, however building antennas that would excite traveling waves in a magnet bore or a TL is much easier. Lifting the limitation of a cutoff frequency of traveling wave NMR, as shown in this work, provides further freedom of implementation in a variety of geometries and for multinuclear operation, or for measuring several or extended samples along a TL.

**References**

(1) Brunner, D. O.; De Zanche, N.; Frohlich, J.; Paska, J.; Pruessmann, K. P. *Nature* **2009**, *457*, 994-992.

(2) Brunner, D. O.; Paska, J.; Froehlich, J.; Pruessmann, K. P. In *Proceedings of the Seventeenth Scientific Meeting of the International Society for Magnetic Resonance in Medicine* Honolulu, HI, USA, 2009; Vol. 499.

(3) Hoffmann, J.; Shajan, G.; Pohmann, R. In *Proceedings of the Eighteenth Scientific Meeting of the International Society for Magnetic Resonance in Medicine* Stockholm, Sweden, 2010; Vol. 3802.

(4) Wiggins, G. C.; Zhang, B.; Duan, Q.; Sodickson, D. K. In *Proceedings of the International Society for Magnetic Resonance in Medicine* Honolulu, HI, USA, 2009; Vol. 2942.

(5) Alt, S.; Müller, M.; Umathum, R.; Bock, M. In *Proceedings of the International Society for Magnetic Resonance in Medicine* Stockholm, Sweden, 2010; Vol. 3795.

(6) Andreychenko, A.; Klomp, D. W.; van den Bergen, B.; van de Bank, B. L.; Kroeze, H.; Lagendijk, J. J.; Luijten, P. R.; van den Berg, C. A. In *Proceedings of the International Society for Magnetic Resonance in Medicine* Honolulu, HI, USA, 2009; Vol. 500.

(7) Kroeze, H.; Andreychenko, A.; van den Berg, C. A.; Klomp, D. W.; Luiten, P. R. In *Proceedings of the Eighteenth Scientific Meeting of the International Society for Magnetic Resonance in Medicine* Stockholm, Sweden, 2010; Vol. 3791.

(8) Zhang, B.; Wiggins, G. C.; Duan, Q.; Lattanzi, R.; Sodickson, D. K. In *Proceedings of the International Society for Magnetic Resonance in Medicine* Honolulu, HI, USA, 2009; Vol. 498.

(9) Kao, C.-P.; Cao, Z.; Oh, S.; Ryu, Y. C.; Collins, C. M. In *Proceedings of the Eighteenth Scientific Meeting of the International Society for Magnetic Resonance in Medicine* Stockholm, Sweden, 2010; Vol. 1446.

(10) Kroeze, H.; van de Bank, B. L.; Visser, F.; Lagendijk, J.; Luijten, P.; Klomp, D. W.; van den Berg, C. In *Proceedings of the Seventeenth Scientific Meeting of the International Society for Magnetic Resonance in Medicine* Honolulu, HI, USA, 2009; Vol. 1320.

(11) Webb, A. G.; Collins, C. M.; Versluis, M. J.; Kan, H. E.; Smith, N. B. *Magn. Res. Med.* **2010**, *63*, 297-302.

(12) Wiggins, G.; Zhang, B.; Lattanzi, R.; Sodickson, D. K. In *Proceedings of the Eighteenth Scientific Meeting of the International Society for Magnetic Resonance in Medicine* Stockholm, Sweden, 2010; Vol. 429.

(13) Zhang, B.; Wiggins, G. C.; Duan, Q.; Sodickson, D. K. In *Proceedings of the Seventeenth Scientific Meeting of the International Society for Magnetic Resonance in Medicine* Honolulu, HI, USA, 2009; Vol. 4746.

(14) van den Berg, C.; Kroeze, H.; van de Bank, B.; van den Bergen, B.; Luijten, P.; Lagendijk, J.; D, K. In *Proceedings of the Seventeenth Scientific Meeting of the International Society for Magnetic Resonance in Medicine* Honolulu, HI, USA, 2009; Vol. 2944.

(15) Jackson, J. *Classical Electrodynamics*; Academic Press: New York, 1998.

(16) Cross, V. R.; Hester, R. K.; Waugh, J. S. *Rev. Sci. Instrum.* **1976**, *47*, 1486-1488.

(17) Gordon, R. E.; Timms, W. E. *J. Magn. Reson.* **1982**, *46*, 322-324.

(18) Rath, A. R. *Magn. Res. Med.* **1990**, *13*, 370-377.

(19) Vaughan, J. T.; Hetherington, H. P.; Otu, J. O.; Pan, J. W.; Pohost, G. M. *Magn. Res. Med.* **1994**, *32*, 206-218.

(20) Walton, J. H.; Conradi, M. S. *J. Magn. Reson.* **1989**, *81*, 623-627.

(21) Maclaughlin, D. E. *Rev. Sci. Instrum.* **1989**, *60*, 3242-3248.

(22) Bodganov, G.; Ludwig, R. *Concepts Magn. Reson.* **2003**, *16B*, 22-37.

(23) Durr, W.; Rauch, S. *Magn. Res. Med.* **1991**, *19*, 446-455.

(24) Matsuzawa, H.; Nakada, T. *Magn. Reson. Mat. Phys. Biol. Med.* **1996**, *4*, 3-6.

(25) van Vaals, J. J.; Bergman, A. H. *J. Magn. Reson.* **1990**, *89*, 331-342.

(26) Gerald, R. E.; Klingler, R. J.; Sandi, G.; Johnson, C. S.; Scanlon, L. G.; Rathke, J. W. *J. Power Sources* **2000**, *89*, 237-243.

(27) Gerald, R. E.; Sanchez, J.; Johnson, C. S.; Klingler, R. J.; Rathke, J. W. *J. Phys.-Condes. Matter* **2001**, *13*, 8269-8285.

(28) Woelk, K.; Gerald, R. E.; Klingler, R. J.; Rathke, J. W. *J. Magn. Reson. Ser. A* **1996**, *121*, 74-77.

(29) Schneider, H. J.; Dullenkopf, P. *Rev. Sci. Instrum.* **1977**, *48*, 832-834.

(30) Schneider, H. J.; Dullenkopf, P. *Rev. Sci. Instrum.* **1977**, *48*, 68-73.

(31) Hull, L. A. *J. Chem. Educ.* **1990**, *67*, 782-783.

(32) Liang, Z.-P.; Lauterbur, P. C. *Principles of magnetic resonance imaging: A signal processing perspective*; Wiley, John & Sons, Inc: New York, 2000.

(33) Zhang, X.; Wang, C.; Vigneron, D. B.; Nelson, S. J. In *Proceedings of the Seventeenth Scientific Meeting of the International Society for Magnetic Resonance in Medicine* Honolulu, HI, USA, 2009; Vol. 104.

(34) Zhang, X.; Wang, C.; Xie, Z.; Wu, B. In *Proceedings of the Sixteenth Scientific Meeting of the International Society for Magnetic Resonance in Medicine* Toronto, ON, Canada, 2008; Vol. 435.

(35) Bobroff, S.; McCarthy, M. J. *Magn. Reson. Imaging* **1999**, *17*, 783-789.

(36) Zhang, J.; Balcom, B. J. *Magn. Reson. Imaging* **2010**, *28*, 826-833.

**Figure Captions.**

**Figure 1.** Transmission line (TL) setup shown with a zoomed in region of the sample holder in line with the coaxial cable inset. The TL runs through the center of the bore of the magnet and the sample is centered in the superconducting coil.

**Figure 2.** $^1$H NMR spectra of mineral oil using (a) a 4.0 mm MAS NMR probe and (b) a TL terminated with a high power attenuator.

**Figure 3.** Experimental (dots) and simulated (line) flip angle distributions of the TL. The $B_1$ field decreases radially outward with inverse distance $r$ from the center, and the relative weighting increases with $r^2$ due to the change in circumference at this distance.

**Figure 4.** (a) The lengths of the cable before and after the sample in the TL setup, where $L = 21$ $\lambda/2$ ($x = 18 \lambda/2$ and $y = 3 \lambda/2$). (b, c) Integrated intensity of the $^1$H NMR signal obtained after adding a length $l$ of cable ranging from $\lambda/2$ to $3\lambda/2$ in increments of $\lambda/8$ (*x*-axis), either (b) before ($x+l$) or (c) after ($y+l$) the sample. The experiments were conducted with open (circles), 50 Ohm matched (square) and short (triangle) circuits.

**Figure 5.** (a) $^1$H NMR Projected spectra and (b) 2D images of mineral oil with attenuated, open, and short circuited TL. A $\lambda/2$ extension cable was added after the sample corresponding to the highest signal intensity with an open circuit.

**Figure 6.** (a) Implementation for multinuclear operation with multiple channels connected to a single TL containing the sample. (b) $^{1}$H-$^{31}$P HMQC NMR spectra of diethyl phosphite (inset) using 4mm MAS probe and TL double resonance attenuated circuit. The magnitude of the TL spectrum was increased by a factor of 5.

**Figure 7.** Possible implementation of cutoff-free TWNMR in a stripline TL.

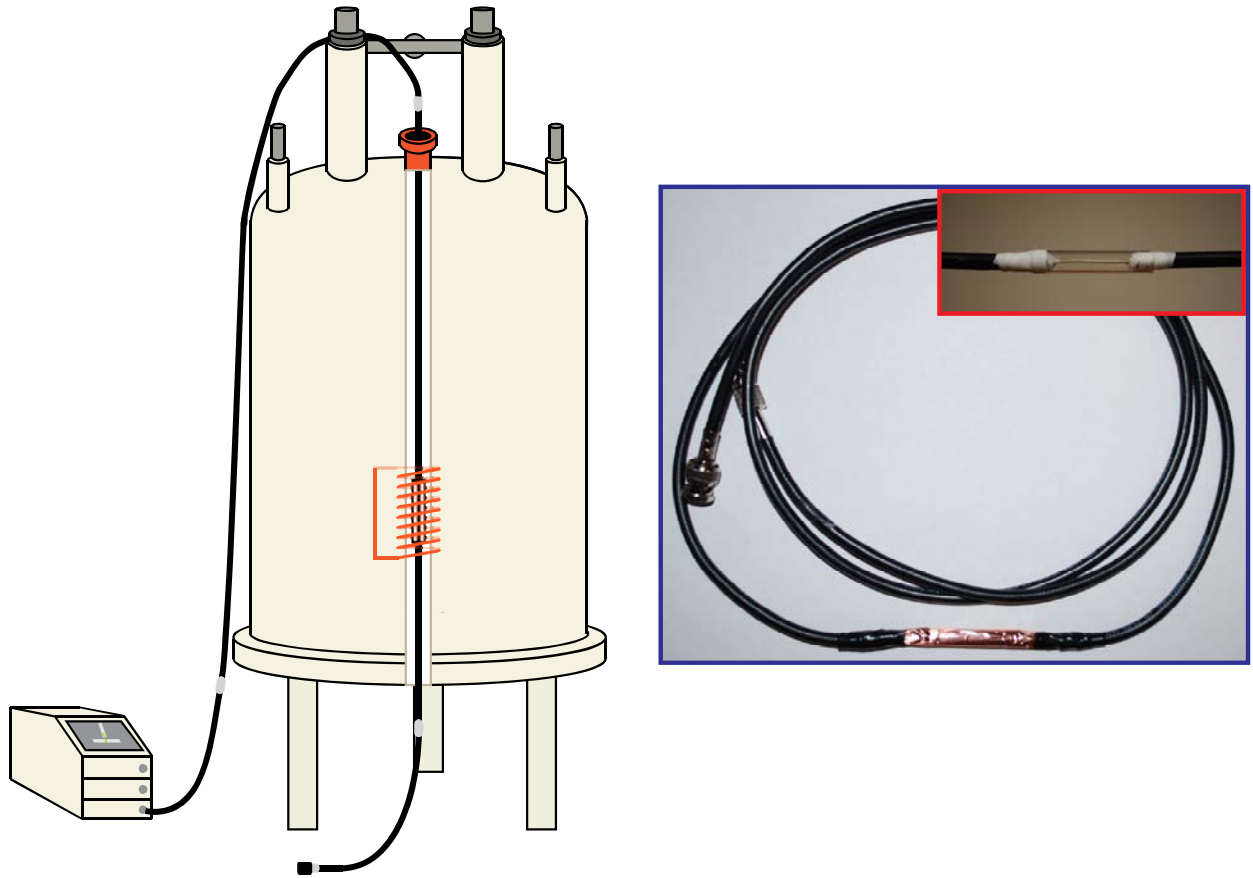

**Figure 1**

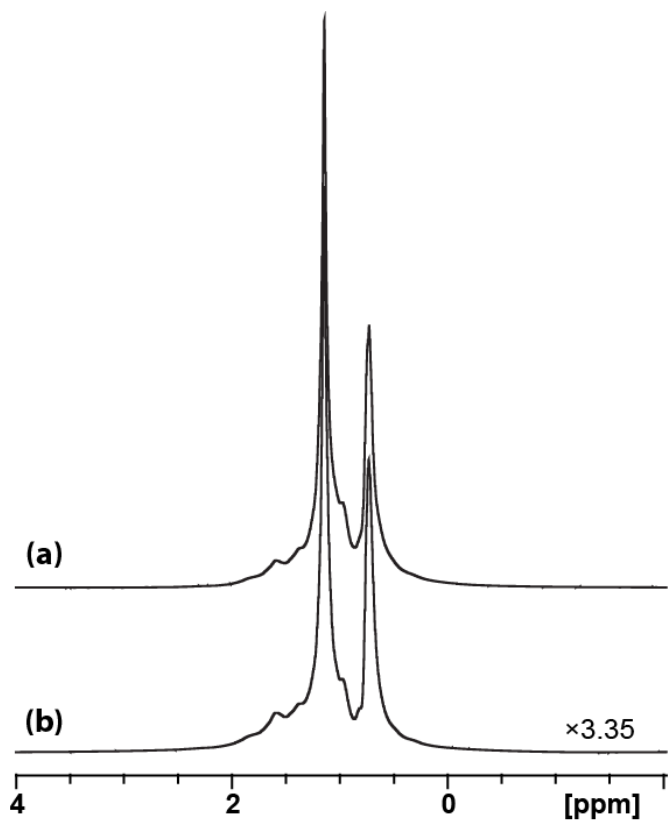

**Figure 2**

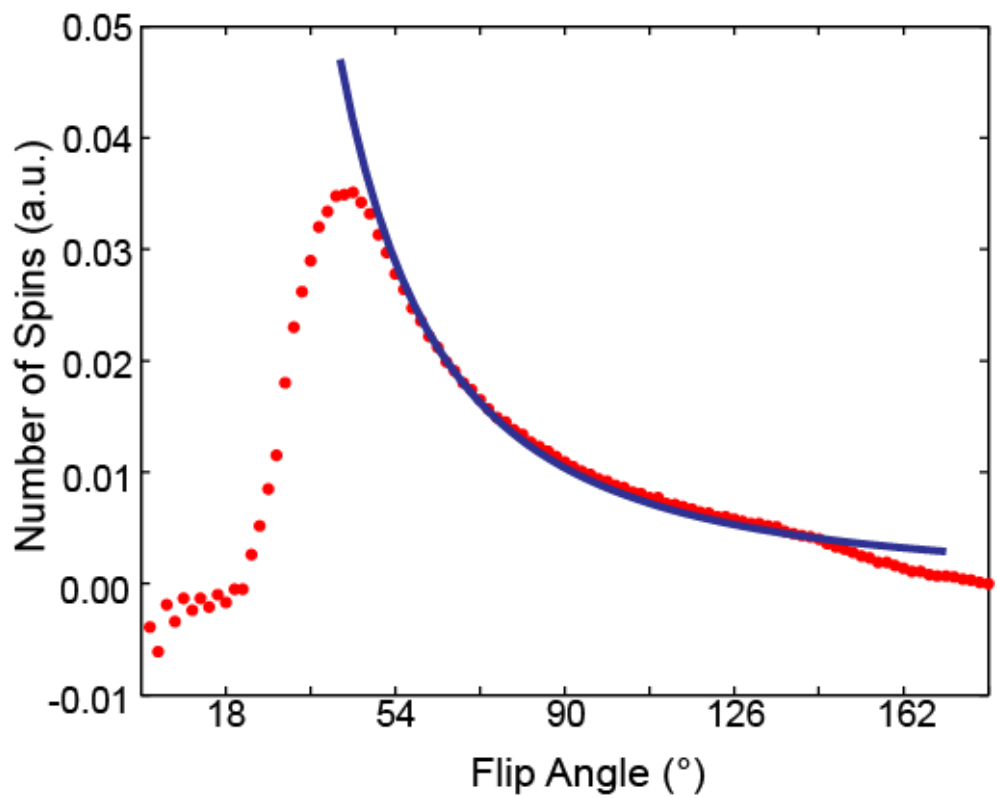

**Figure 3**

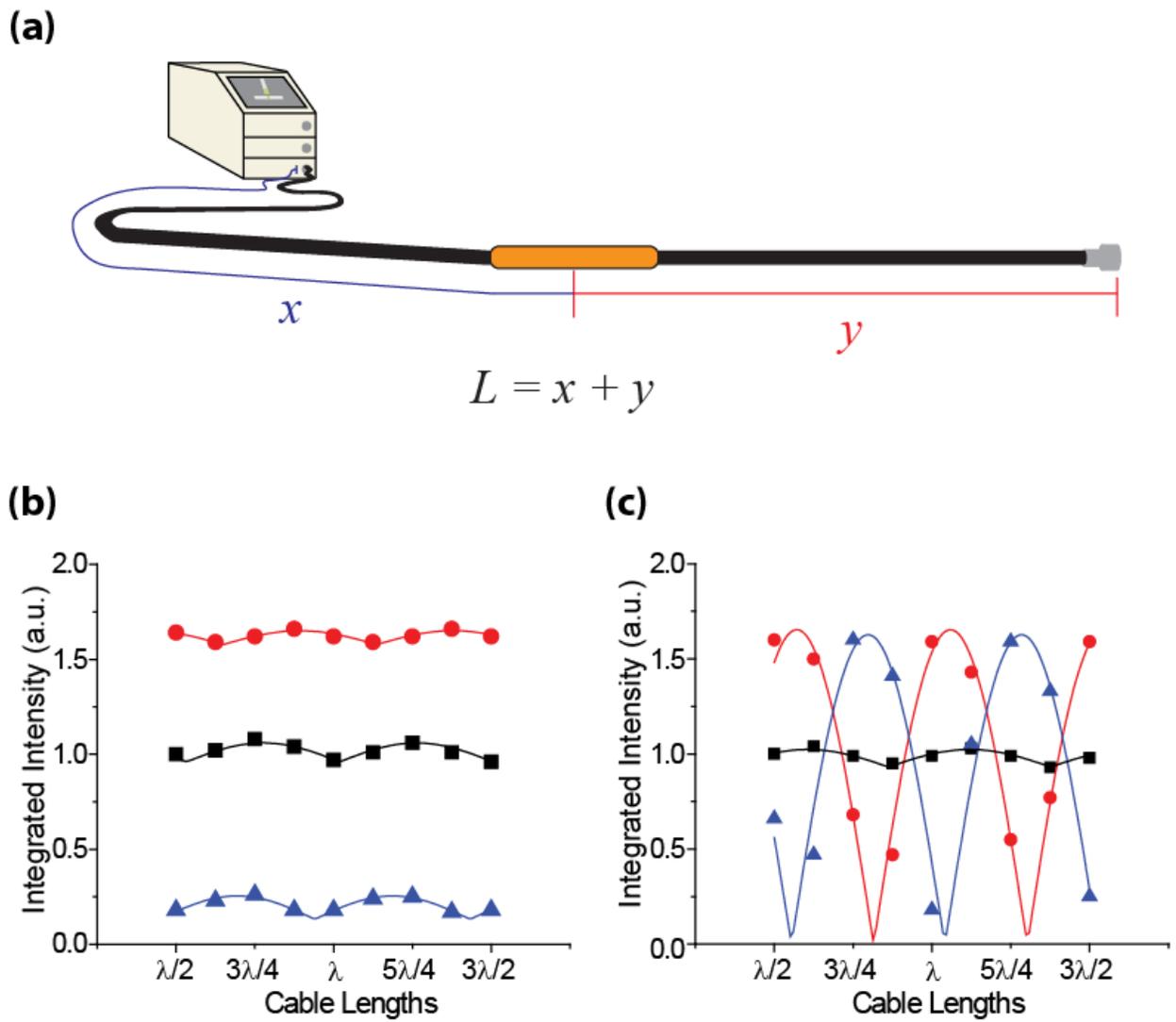

**Figure 4**

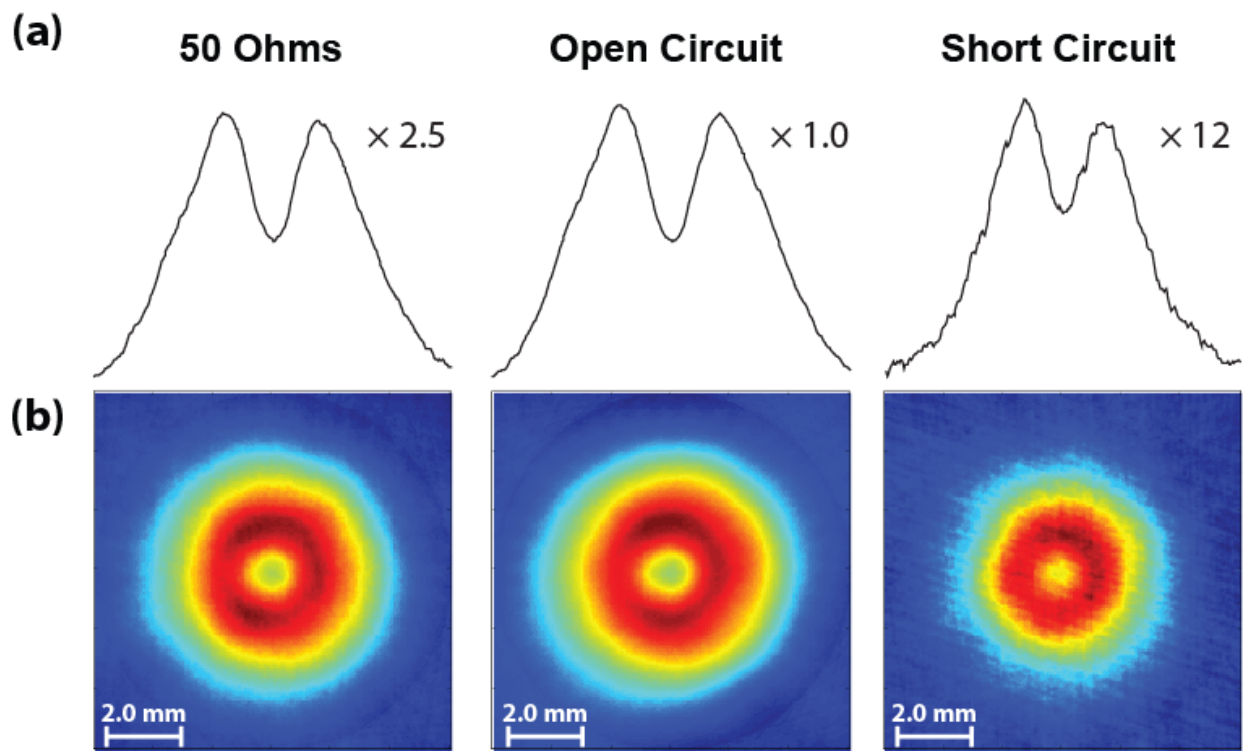

**Figure 5**

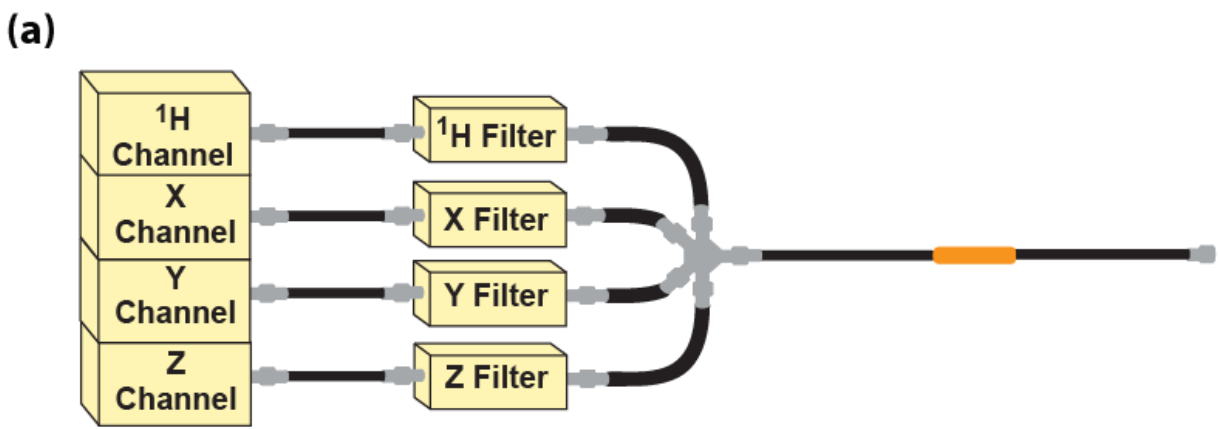

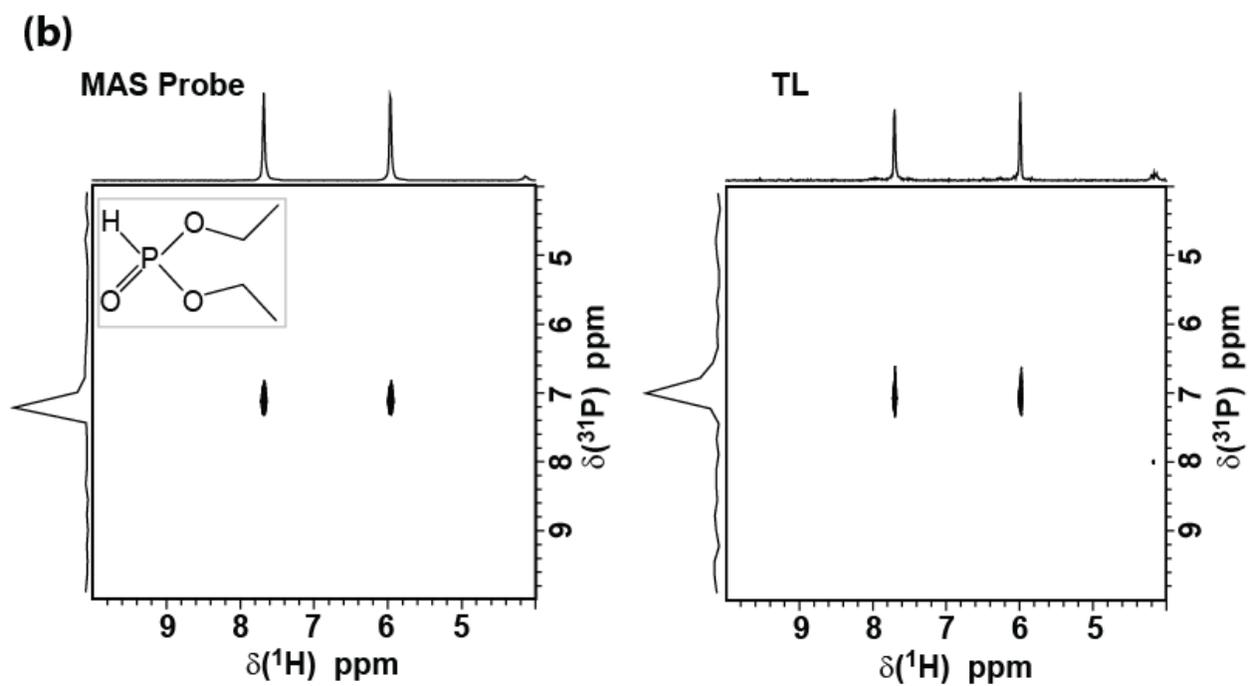

**Figure 6**

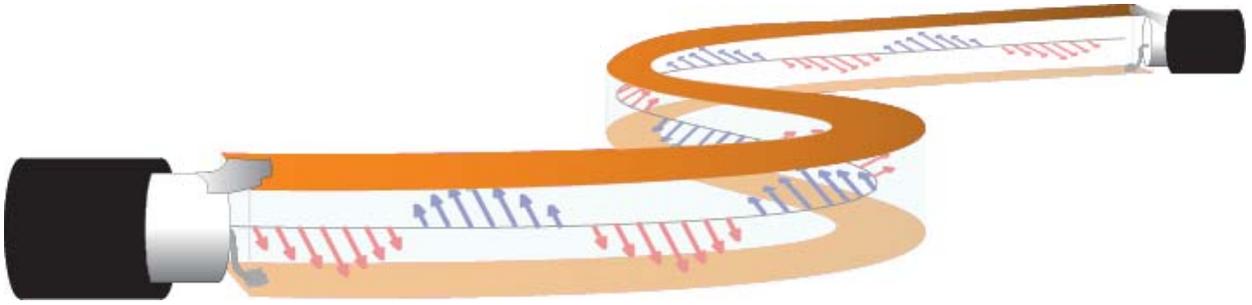

**Figure 7**